\documentclass[12pt,a4paper,usenatbib]{article}
\usepackage{graphicx}
\usepackage{times}
\usepackage{amsmath}
\usepackage{amssymb}
\usepackage{xcolor}
\usepackage{enumitem}
\usepackage{subfig}
\usepackage[para]{threeparttable}
\usepackage{footmisc}
\usepackage{hyperref}

\title{\bf DustKING - the story continues:\\ dust attenuation in NGC\,628}

\author{Marjorie Decleir$^1$\thanks{E-mail: Marjorie.Decleir@UGent.be}, Ilse De Looze$^{1,2}$,
M\'ed\'eric Boquien$^3$, Maarten Baes$^1$\\
\vspace{0.5cm}\\
\normalsize $^1$ Sterrenkundig Observatorium, Universiteit Gent, Krijgslaan 281 S9, 9000 Gent, Belgium\\
\normalsize $^2$ Dept. of Physics \& Astronomy, University College London, Gower Street, London WC1E 6BT, UK\\
\normalsize $^3$ Univ. de Antofagasta, Centro de Astronom\'ia, Avenida Angamos 601, Antofagasta 1270300, Chile}
\date{\mbox{}}
\begin{document}
\maketitle
\setcounter{page}{1001}
\pagestyle{plain}
    \makeatletter
    \renewcommand*{\pagenumbering}[1]{%
       \gdef\thepage{\csname @#1\endcsname\c@page}%
    }
    \makeatother
\pagenumbering{arabic}

%
% WE REDEFINE THE plain LaTeX PAGESTYLE !!! 
% THIS PAGESTYLE WILL BE USED FOR THE FIRST PAGE ONLY !
% Please do not change the following lines
%
\def\bull{\vrule height .9ex width .8ex depth -.1ex}
\makeatletter
\def\ps@plain{\let\@mkboth\gobbletwo
\def\@oddhead{}\def\@oddfoot{\hfil\scriptsize\bull\quad
"2nd Belgo-Indian Network for Astronomy \& astrophysics (BINA) workshop'', held in Brussels (Belgium), 9-12 October 2018 \quad\bull}%
\def\@evenhead{}\let\@evenfoot\@oddfoot}
\makeatother
%
% AND DEFINE OUR MACROS FOR THE REFERENCE LIST
% I.E \beginrefer \refer and \endrefer
%
\def\beginrefer{\section*{References}%
\begin{quotation}\mbox{}\par}
\def\refer#1\par{{\setlength{\parindent}{-\leftmargin}\indent#1\par}}
\def\endrefer{\end{quotation}}
%
% BEGIN THE ABSTRACT WITH \noindent\small, ENCLOSE IT IN A GROUP
% AND BOLDFACE THE TITLE.
%
{\noindent\small{\bf Abstract:}
Dust attenuation is a crucial but highly uncertain parameter that hampers the determination of intrinsic galaxy properties, such as stellar masses, star formation rates and star formation histories. The shape of the dust attenuation law is not expected to be uniform between galaxies, nor within a galaxy. Our DustKING project was introduced at the first BINA workshop in 2016 and aims to study the variations of dust attenuation curves in nearby galaxies. At the second BINA workshop in 2018, I presented the results of our pilot study for the spiral galaxy NGC\,628. We find that the average attenuation law of this galaxy is characterised by a MW-like bump and a steep UV slope. Furthermore, we observe intriguing variations within the galaxy, with regions of high $A_{\text{V}}$ exhibiting a shallower attenuation curve. Finally, we discuss how our work might benefit from data taken with the UVIT from the Indian AstroSat mission.}
\vspace{0.5cm}\\
% SPECIFY UP TO 5 KEYWORDS SEPARATED BY ' -- '
\section{Introduction}

Although interstellar dust only makes up a very small fraction of the interstellar mass in galaxies (typically about 1\% or less, e.g. R\'emy-Ruyer et al. 2014), it is a crucial component in galaxies. Dust particles, for example, regulate the heating of the interstellar gas and catalyze the formation of molecular hydrogen, which will form more easily on the surface of a dust grain. Also, the dust will shield these newly formed molecules from the hard ultraviolet (UV) radiation of young stars. These conditions provide a suitable environment for star (and planet) formation in the molecular clouds. Finally, dust grains heavily influence our view on other galaxy components, since they absorb and scatter about 30\% of the starlight (UV and optical radiation) in the Universe (Lagache et al. 2005; Viaene et al. 2016). To obtain an undistorted view of galaxies, understanding the interplay between dust and starlight is essential.

It is important to stress the difference between dust extinction and attenuation, since both terms are often confused. Extinction is the combination of absorption and scattering of light by dust, in which case the light is originating from a background source and all dust is located between the source and the observer as a ``screen" of dust. In this configuration, e.g. when looking at individual stars in the Milky Way (MW), in M31 or in the Magellanic Clouds, the distribution of dust does not affect the total amount of extinction measured by the observer and an intrinsic extinction curve can be obtained, giving the amount of dust extinction as a function of wavelength. However, for more distant galaxies, individual stars cannot be resolved and consequently, there is a mix of stars and dust along each sightline towards the observed galaxy. The effect of the dust in this case is called attenuation and can be described by a dust attenuation curve, which is influenced by a combination of intrinsic dust properties and geometry effects. The specific shape of a dust extinction/attenuation law is characterized by the steepness of its UV slope, and the strength of the 2175\,\AA\ bump (i.e. an excess in dust absorption around that wavelength).

Since the knowledge about the dust properties outside our own Galaxy and the Magellanic Clouds is still rather limited, it is usually assumed that the dust composition in other galaxies is the same as observed in the MW. However, there is growing evidence for strong deviations from a universal dust attenuation law, based on the different shapes of dust attenuation curves observed in other galaxies (e.g. Salim et al. 2018, Battisti et al. 2017a,b; Reddy et al. 2015; Salmon et al. 2016).  Dust attenuation is still one of the most uncertain parameters to recover intrinsic galaxy properties such as the masses of the stellar populations, the star formation rates (SFRs) and the star formation histories (SFHs). To overcome this problem, we launched the DustKING project in which we investigate the variations in the shape of dust attenuation curves and dust properties for the KINGFISH (Key Insights on Nearby Galaxies: a Far-Infrared Survey with Herschel, Kennicutt et al. 2011) sample of nearby galaxies, spanning a wide variety of morphological galaxy types, metallicities and star formation activities.

\section{SWIFT UVOT data}
\label{sec: Data}
\begin{figure}[h]
\centering
\includegraphics[width=10cm]{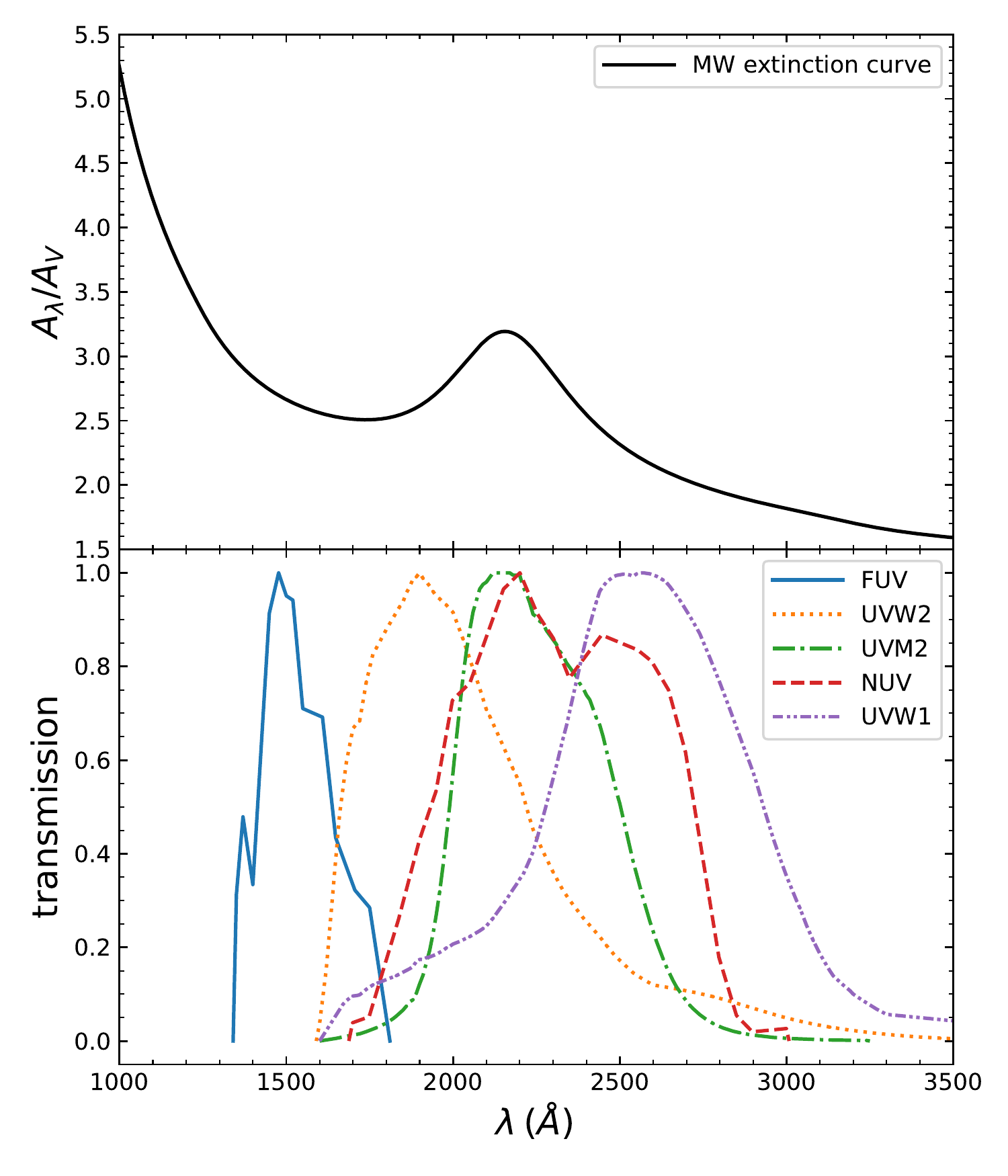}
\caption{Transmission curves of the SWIFT UVOT and the GALEX filters, taken from the SVO Filter Profile Service (\url{http://svo2.cab.inta-csic.es/svo/theory/fps/}). The MW extinction curve in the top panel is taken from Cardelli et al. (1989). Figure taken from Decleir et al. (2019). \label{fig: uvot_filters} }
\end{figure}

In our study, we use a multi-wavelength dataset, ranging from the far-ultraviolet (FUV) to the far-infrared (FIR). In particular, we make use of images taken with the UV/Optical Telescope (UVOT, Roming et al. 2005) onboard the SWIFT satellite (Gehrels et al. 2004). Fig. \ref{fig: uvot_filters} shows the transmission curves of the three UVOT broadband filters (UVW2, UVM2 and UVW1) and the GALEX FUV and NUV filters, together with an average MW extinction curve, with its typical bump feature around 2175\,\AA. It can be seen that the UVOT filters nicely cover this bump, which makes them useful to constrain its amplitude and help us understand what is causing this excess.

Since the SWIFT telescope was designed to detect point sources, the standard SWIFT data reduction pipeline is not suitable for our purposes. Therefore, we developed a new reduction pipeline, entirely optimized for extended sources, which is presented in Decleir et al. (2019) and will be made publicly available in Decleir et al. (in prep.).

\section{Results for the pilot study on NGC\,628}
\label{sec: results}
\subsection{Dust attenuation properties on resolved scales}

\begin{figure}[h]
\centering
	\includegraphics[width=12cm]{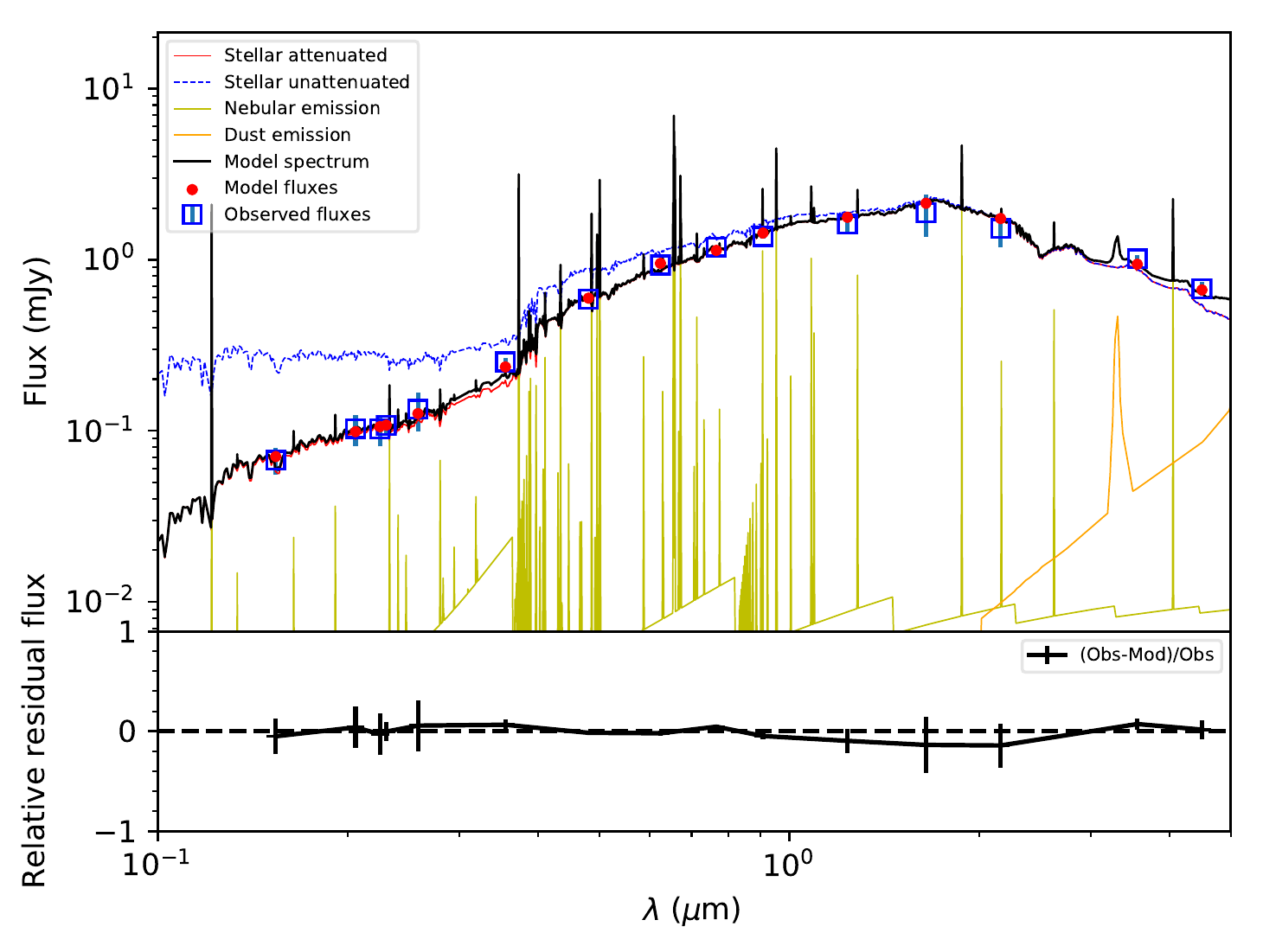}
    \caption{Top: Best-fitting model SED for pixel 261,258 obtained with \textsc{cigale}. Bottom: Relative residual flux between model and observations. \label{fig: SED_highSFR}}
\end{figure}

We did a pilot study on the dust attenuation properties in NGC\,628 (or M74), a nearby grand design spiral galaxy, on spatially resolved scales of about 325\,pc (or 7"). In order to constrain the dust attenuation curve, we fit the observed spectral energy distribution (SED), on a pixel-by-pixel base, with the state-of-the art code \textsc{cigale}\footnote{\url{https://cigale.lam.fr/}} (a \textsc{python} Code Investigating GALaxy Emission, Noll et al. 2009, Roehlly et al. 2014, Boquien et al. 2018). \textsc{cigale} is based on the dust energy balance principle: the energy that was absorbed by dust in the UV-optical range, will be re-emitted by the dust at IR wavelengths. In Fig. \ref{fig: SED_highSFR} an example of a fitted SED can be found. Overall, the observed data points are fitted very well.

\begin{figure}
	\includegraphics[width=\textwidth]{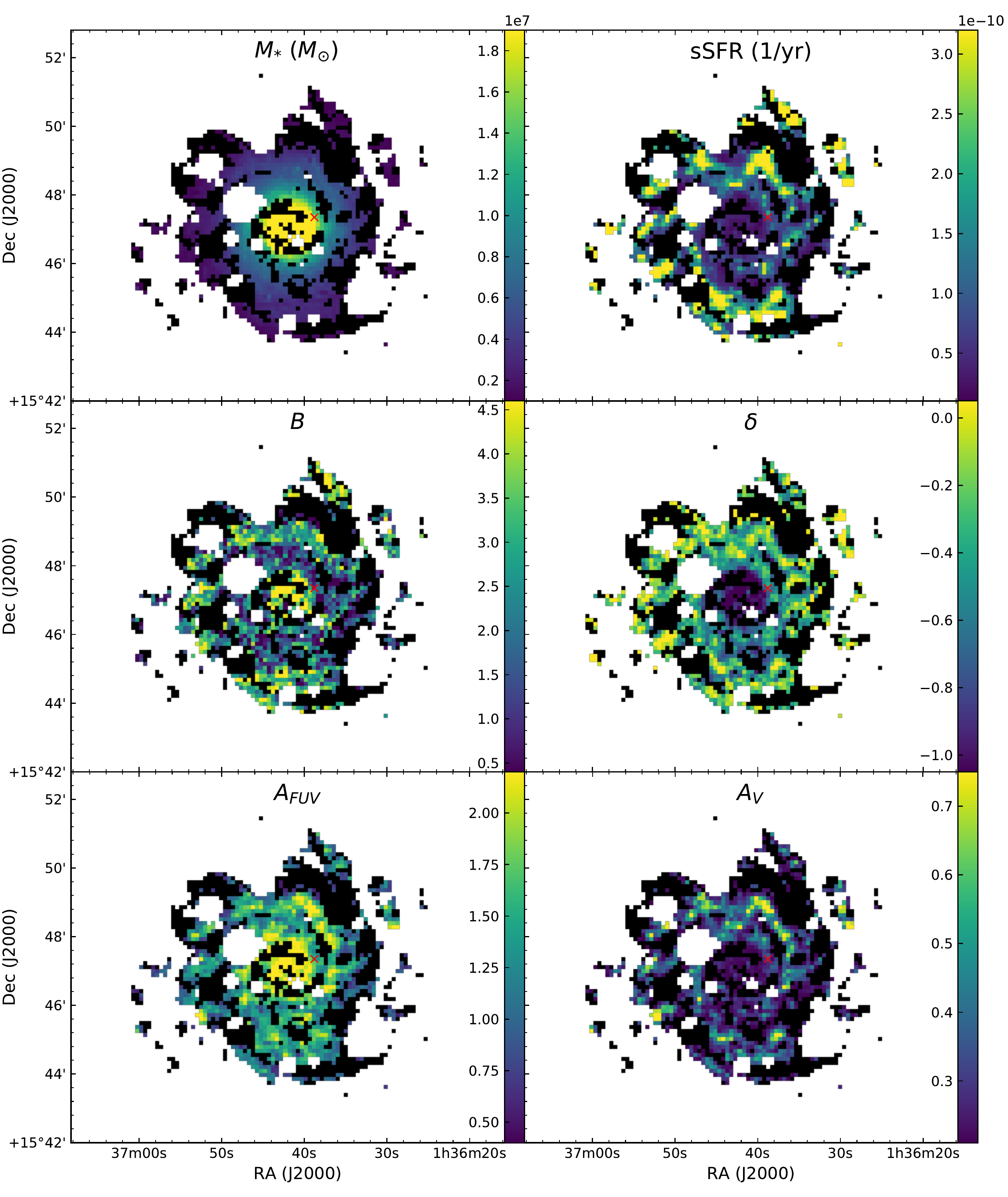}
    \caption{Maps of the stellar mass $M_{*}$ (in $M_{\odot}$), the sSFR (in 1/yr), the UV bump strength $B$, the attenuation slope $\delta$, the FUV attenuation $A_{\text{FUV}}$ and the V-band attenuation  $A_{\text{V}}$ for NGC\,628, all derived from the \textsc{cigale} fitting. Pixels with $\mathbf{A_\textrm{V}\leq0.2}$ are shown in black. The red cross indicates pixel 261,258 in the central spiral arm, for which the SED can be found in Fig. \ref{fig: SED_highSFR}.}
    \label{fig: maps}
\end{figure}

The SED fitting was done for every pixel in the galaxy image (with reduced $\chi^2$ between 1 and 6, and a median reduced $\chi^2 = 3.28$), which allows us to create maps of the different galaxy properties. In Fig. \ref{fig: maps} we show the distribution of the parameter values obtained with \textsc{cigale} across the disc of NGC\,628. The stellar mass ($M_{*}$) map shows a smooth distribution of the stars, with most stars located in the centre of the galaxy. The specific SFR map reveals the location of the star forming regions, mostly in the spiral arms of the galaxy. The maps of the bump strength $B$ and the slope $\delta$ show the spatial variations in the dust attenuation curve within the galaxy.

\subsection{Trends between the shape of the dust attenuation curve and other galaxy properties}
\label{sec: Trends}

\begin{figure}[h]
\centering
\includegraphics[width=10 cm]{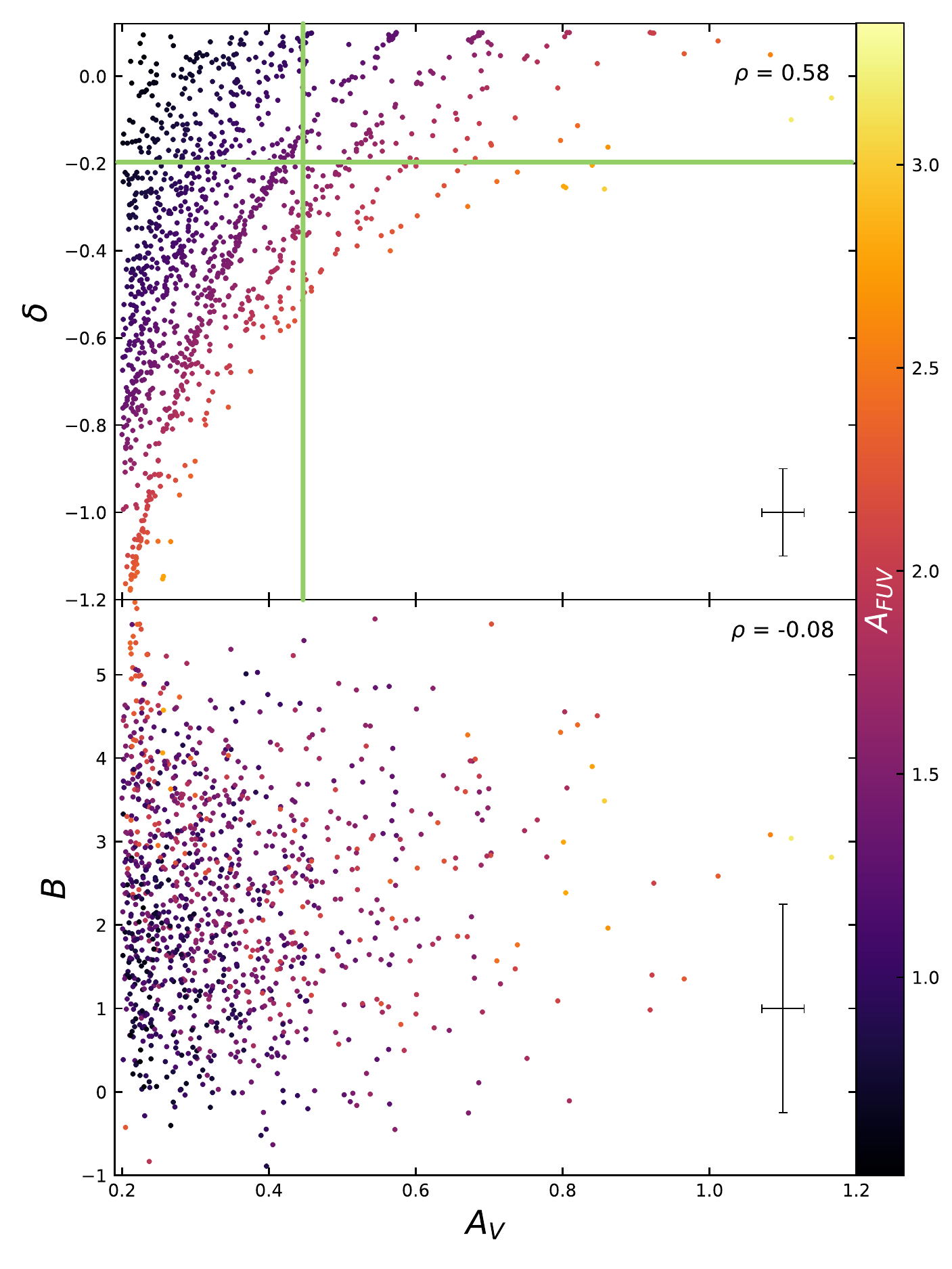}
\caption{The attenuation curve slope $\delta$ and the bump strength $B$ vs. the V-band attenuation $A_{\text{V}}$, colour-coded with the FUV attenuation $A_{\text{FUV}}$. Typical error bars as well as the Spearman correlation coefficient $\rho$ are indicated. \label{fig: AV}}
\end{figure}

It is interesting to verify whether the variations in bump strengths and slopes correlate with other galaxy properties. In the top panel of Fig. \ref{fig: AV}, we observe a trend between the slope of the attenuation curve and the V-band attenuation $A_{\text{V}}$: in galaxy regions with a higher attenuation level, the slope of the curve is shallower. A similar trend was found by Hagen et al. (2017), Salim et al. (2018), Tress et al. (2018) and Narayanan et al. (2018), and interpreted in light of models presented in Chevallard et al. (2013). This trend can be explained by geometry effects: in regions with higher levels of dust attenuation, dust absorption dominates over scattering processes which results in a flattening of the dust attenuation curve. This suggests that the relative geometry of stars and dust dominates variations in the shape of the dust attenuation curve as opposed to alterations in intrinsic grain species.

In the same plot, the colour scale reveals a trend between the slope and the FUV attenuation: pixels with a higher $A_{\text{FUV}}$ seem to have steeper slopes. This was also found by Salim et al. (2018) and may appear contradictory to the slope vs. $A_{\text{V}}$ trend, because one would expect that the dust attenuation in the UV and in the optical are linked and at least scale in the same way. However, as clearly explained in their work, this is only true if the galaxy obeys a universal attenuation law. Looking at a fixed $A_{\text{V}}$ in Fig. \ref{fig: AV} (i.e. following a vertical line), shows that a shallower slope corresponds to lower $A_{\text{FUV}}$ values. This can be understood from the attenuation curves in Fig. \ref{fig: att_curves} which are all normalized to $A_{\text{V}}$: for a fixed $A_{\text{V}}$, shallower slopes result in a lower attenuation in the FUV, compared to attenuation curves with steeper slopes. Or, one can look at it in a different way: for a fixed slope (i.e. following a horizontal line in Fig. \ref{fig: AV}), a higher $A_{\text{V}}$ indeed corresponds to a higher $A_{\text{FUV}}$. In other words, if the slope is fixed, the curve will simply scale up and down if there is more or less attenuation.

The bottom panel of Fig. \ref{fig: AV} shows no correlation between the bump strength and $A_{\text{V}}$ (nor $A_{\text{FUV}}$), in contrast to Hagen et al. (2017), Salim et al. (2018) and Tress et al. (2018) who found weaker bumps at higher $A_{\text{V}}$. It is possible that in our case the somewhat larger uncertainty on the derived bump strengths masks such a correlation.

\subsection{Comparison with other extinction/attenuation curves}

\begin{figure}[h]
\centering
\includegraphics[width=12cm]{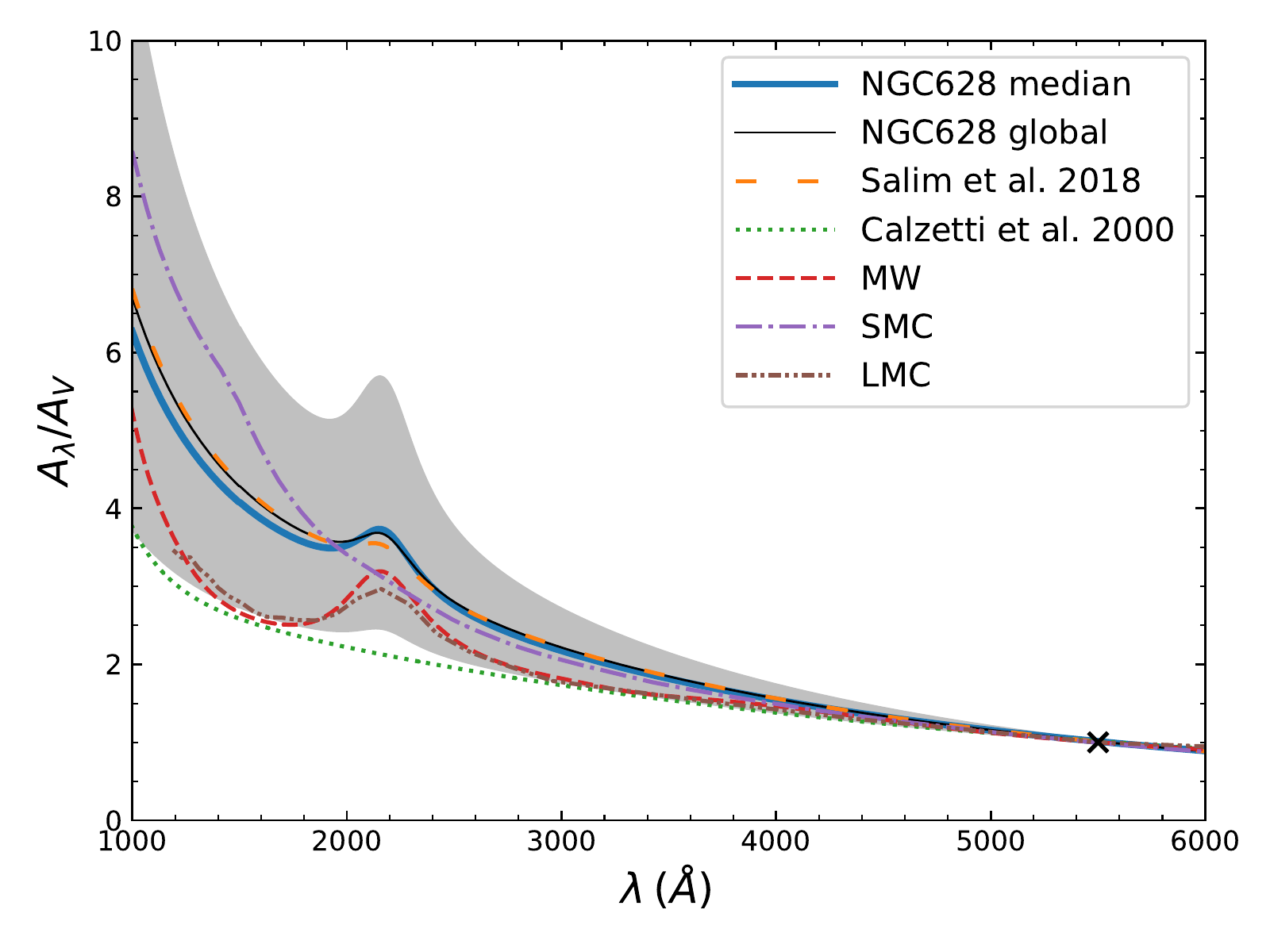}
\caption{Extinction/attenuation curves for different galaxies. See text for more details. \label{fig: att_curves}}
\end{figure}

For comparison, we also calculated a ``median" attenuation curve for NGC\,628, characterized by the median bump and the median slope. It is shown in Fig. \ref{fig: att_curves}, together with some other relevant curves: the MW extinction curve (from Cardelli et al. 1989), the SMC and LMC extinction curves (from Gordon et al. 2003), the Calzetti et al. (2000) curve (which is often used as the standard attenuation law for starburst galaxies), and the average curve found by Salim et al. (2018). It is clear that the median NGC\,628 curve is a lot steeper than the MW and Calzetti curves, but not as steep as the SMC curve, and has a bump that is somewhat smaller, but consistent within the error bars, compared to the MW curve. The shaded region in the plot represents the attenuation curves between the 16\textsuperscript{th} and 84\textsuperscript{th} percentile of bump and slope values. We refer the reader to Decleir et al. (2019) for an in-depth discussion and more details on these results.

\section{Conclusions and future outlook}
\label{sec: Summary}
In this work, we present our study on the dust attenuation properties on resolved scales in the nearby spiral galaxy NGC\,628 using SWIFT UVOT observations that probe the 2175\,\AA\ dust absorption feature. The main conclusions of our analysis can be summarized as follows:

\begin{itemize}
    \item NGC\,628 has a steep attenuation curve (with a slope in between that of the MW and the SMC curves), with a prominent bump (consistent with the MW curve).
    \item Variations in the shape of the attenuation curve can be observed within the galaxy: shallower slopes are found in regions with higher levels of dust attenuation, which can be explained by geometry effects.
\end{itemize}

To understand how the dust attenuation properties vary in the nearby Universe, we require a study of the dust attenuation curve in a statistical sample of nearby galaxies, for which the modelling has been done in a homogeneous way. To this aim, we are currently extending the analysis presented here to a larger sample of nearby galaxies (Decleir et al., in prep.).

Since the attenuation curve mainly varies in the UV-wavelength range, UV data is essential for this kind of study. As stated above, we are using data from both the GALEX and the SWIFT missions to cover the UV. However, as can be seen in Fig. \ref{fig: uvot_filters}, the UVOT filters are very broad and the UVW2 and UVW1 filters have long tails overlapping with the bump. As a consequence, these bands do not uniquely measure the slope of the attenuation curve, but are also affected by its bump.

\begin{figure}[h]
\centering
\includegraphics[width=10cm]{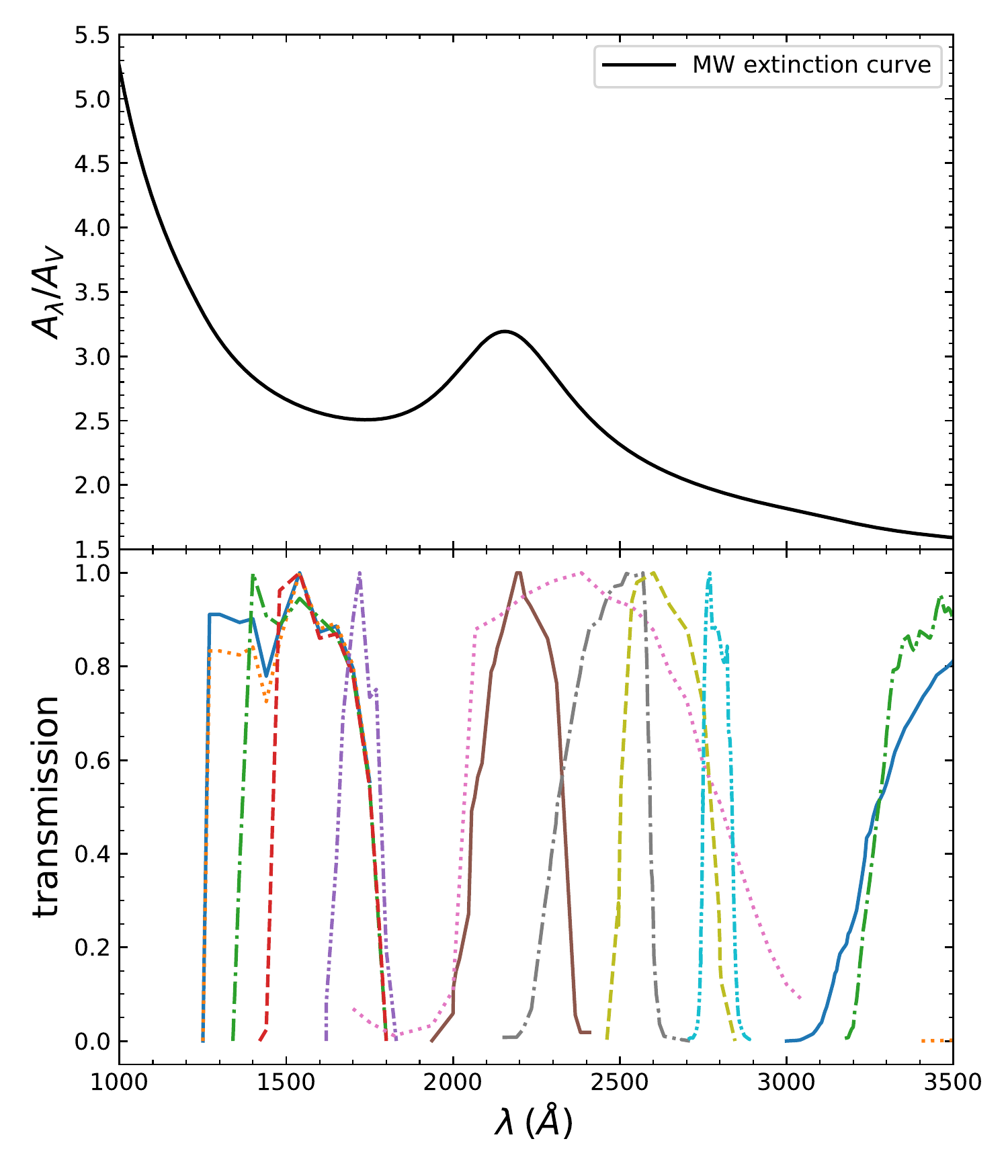}
\caption{Transmission curves of the AstroSat UVIT filters, taken from the SVO Filter Profile Service (\url{http://svo2.cab.inta-csic.es/svo/theory/fps/}). \label{fig: UVIT}}
\end{figure}

In Fig. \ref{fig: UVIT} the transmission curves of the filters on the AstroSat UltraViolet Imaging Telescope (UVIT, Hutchings 2014) are shown. UVIT has more filters than SWIFT and covers a larger wavelength range. Furthermore, the filters are narrower, and some of them are not influenced by the bump. Therefore, we expect that UVIT data will greatly assist in constraining the bump strength and the slope of the attenuation curve. We plan to use archival UVIT data, and to submit a proposal to observe a sample of nearby galaxies with UVIT in the near future.

%
% USE A SECTION WITHOUT NUMBER FOR THE ACKNOWLEDGEMENTS
%
\section*{Acknowledgements}
This work benefits from a PhD Fellowship of the Research Foundation - Flanders (FWO-Vlaanderen).
I. De Looze gratefully acknowledges the support of the Research Foundation - Flanders. M. Boquien was supported by MINEDUC-UA projects, code ANT 1655 and ANT 1656, and FONDECYT project 1170618.

We want to thank the organizers of the second BINA workshop for their hospitality.

%
% BEGIN THE REFERENCE LIST WITH \beginrefer
% USE \refer BEFORE THE REFERENCES AND BEGIN A NEW PARAGRAPH AFTER THE 
% REFERENCE !
% DO NOT FORGET TO END THE LIST WITH \endrefer
% 
%
% INSTRUCTIONS FOR BIBLIOGRAPHY:
% ==============================
% - DON'T USE THE & SYMBOL
% - USE INITIALS FOR FIRST AND MIDDLE NAMES, AND SPECIFY FULL FAMILY NAME (see examples below)
% - NO COMMA BETWEEN NAME AND INITIALS
% - USE COMMA BETWEEN DIFFERENT AUTHORS NAMES
% - NO COMMA AFTER THE LAST AUTHOR NAME
% - FOR LONG AUTHOR LISTS, SPECIFY THE FIRST 3 AUTHORS FOLLOWED BY 'et al.', WITH NO COMMA BEFORE AND AFTER 'et al.'
% - INSERT A BLANK SPACE BETWEEN MULTIPLE INITIALS
% - USE STANDARD JOURNAL ACRONYMS FREQUENTLY USED IN MAIN ASTROPHYSICS JOURNAL
% - SORT REFERENCES BY ALPHABETICAL ORDER OF FIRST AUTHOR NAMES

\footnotesize
\beginrefer

\refer Battisti A. J., Calzetti D., Chary R.-R. 2017a, ApJ, 840, 109

\refer Battisti A. J., Calzetti D., Chary R.-R. 2017b, ApJ, 851, 90

\refer Boquien M., Burgarella D., Roehlly Y. et al. 2018, A\&A, 622, A103

\refer Calzetti D., Armus L., Bohlin R. C. et al. 2000, ApJ, 533, 682

\refer Cardelli J. A., Clayton G. C., Mathis J. S. 1989, ApJ, 345, 245

\refer Chevallard J., Charlot S., Wandelt B., Wild V. 2013, MNRAS, 432, 2061

\refer Decleir M., De Looze I., Boquien M. et al. 2019, arXiv, 190306715

\refer Gehrels N., Chincarini G., Giommi P. et al. 2004, ApJ, 611, 1005

\refer Gordon K. D., Clayton G. C., Misselt K. A., Landolt A. U., Wolff M. J. 2003, ApJ, 594, 279

\refer Hagen L. M. Z., Siegel M. H., Hoversten E. A. et al. 2017, MNRAS, 466, 4540

\refer Hutchings J.B. 2014, Ap\&SS, 354, 143

\refer Kennicutt R. C., Calzetti D., Aniano G. et al. 2011, PASP, 123, 1347

\refer Lagache G., Puget J.-L., Dole H. 2005, ARA\&A, 43, 727

\refer Narayanan D., Conroy C., Dav\'e R., Johnson B. D., Popping G. 2018, ApJ, 869, 70

\refer Noll S., Burgarella D., Giovannoli E. et al. 2009, A\&A, 507, 1793

\refer Reddy N. A., Kriek M., Shapley A. E. et al. 2015, ApJ, 806, 259

\refer R\'emy-Ruyer A., Madden S. C., Galliano F. et al. 2014, A\&A, 563, A31

\refer Roehlly Y., Burgarella D., Buat V. et al. 2014, ASPC, 485, 347

\refer Roming P. W. A., Kennedy T. E., Mason K. O. et al. 2005, SSRv, 120, 95

\refer Salim S., Boquien M., Lee J. C. 2018, ApJ, 859, 11

\refer Salmon B., Papovich C., Long J. et al. 2016, ApJ, 827, 20

\refer Tress M., M{\'a}rmol-Queralt{\'o} E., Ferreras I. et al. 2018, MNRAS, 475, 2363

\refer Viaene S., Baes M., Bendo G. et al. 2016, A\&A, 586, A13

\endrefer           

\end{document}